\documentclass[conference]{IEEEtran}
\usepackage{cite}

\ifCLASSINFOpdf
\usepackage[pdftex]{graphicx}
\else
\fi
\usepackage{amsmath}
\usepackage{amsfonts} 
\usepackage{amsthm}
\usepackage{amssymb}
\usepackage[boxruled]{algorithm2e}
\usepackage{url}

\usepackage{bbm}

\usepackage{kbordermatrix}

\newcommand{\pfun}{\mathop{\hbox{$\to$\kern-7pt\raise.9pt\hbox{\scalebox{1}[.55]{$|$}}\kern4pt} }}

\usepackage{array}

\begin{document}

\title{
Fast Mapping onto Census Blocks}

\author{\IEEEauthorblockN{Jeremy Kepner$^1$, Andreas Kipf$^1$, Darren Engwirda$^2$, Navin Vembar$^3$,  Michael Jones$^1$, Lauren Milechin$^1$, \\  Vijay Gadepally$^1$, Chris Hill$^1$, Tim Kraska$^1$,  William Arcand$^1$, David Bestor$^1$, William Bergeron$^1$, \\ Chansup Byun$^1$,  Matthew Hubbell$^1$, Michael Houle$^1$, Andrew Kirby$^1$, Anna Klein$^1$, Julie Mullen$^1$, Andrew Prout$^1$, \\ Albert Reuther$^1$, Antonio Rosa$^1$, Sid Samsi$^1$, Charles Yee$^1$, and Peter Michaleas$^1$
\\
\IEEEauthorblockA{$^1$MIT, $^2$Columbia University, $^3$Camber Systems
}}}
\maketitle

\begin{abstract}
Pandemic measures such as social distancing and contact tracing can be enhanced by rapidly integrating dynamic location data and demographic data.  Projecting billions of longitude and latitude locations onto hundreds of thousands of highly irregular demographic census block polygons is computationally challenging in both research and deployment contexts.  This paper describes two approaches labeled ``simple'' and ``fast''.  The simple approach can be implemented in any scripting language (Matlab/Octave, Python, Julia, R) and is easily integrated and customized to a variety of research goals. This simple approach uses a novel combination of hierarchy, sparse bounding boxes, polygon crossing-number, vectorization, and parallel processing to achieve 100,000,000+ projections per second on 100 servers. The simple approach is compact, does not increase data storage requirements, and is applicable to any country or region.  The fast approach exploits the thread, vector, and memory optimizations that are possible using a low-level language (C++) and achieves similar performance on a single server.  This paper details these approaches with the goal of enabling the broader community to quickly integrate location and demographic data.
\end{abstract}

%
\IEEEpeerreviewmaketitle

\section{Introduction}
\let\thefootnote\relax\footnotetext{
Research was sponsored by the United States Air Force Research Laboratory and was accomplished under Cooperative Agreement Number FA8750-19-2-1000. The views and conclusions contained in this document are those of the authors and should not be interpreted as representing the official policies, either expressed or implied, of the United States Air Force or the U.S. Government. The U.S. Government is authorized to reproduce and distribute reprints for Government purposes notwithstanding any copyright notation herein.
}

Pandemics require urgent solutions to a variety of problems \cite{moturi2018panstop}.  Teams must rapidly assemble, address a problem, and broadly communicate their approach.  This paper describes a solution developed by a team of individuals that came together virtually over a period of days to accelerate integrating location and demographic data.  The availability of devices capable of registering and transmitting location data has increased dramatically over the last decade.  The billions of locations produced every day by these devices is a valuable resource for enabling social distancing, contact tracing, and other pandemic measures \cite{eames2003contact,wang2020response}.

Location data can be significantly enriched when integrated with demographic information found in publicly available United States census data, such as population density and age distribution.  Coupling these data requires rapidly projecting longitude and latitude locations onto hundreds of thousands of highly irregular census block polygons.  The point-in-polygon join problem has been extensively studied (see \cite{kipf2020adaptive} and references therein \cite{DBLP:conf/sigmod/Orenstein89, brinkhoff1994multi, zimbrao1998raster, jacox2007spatial, kothuri2001efficient, DBLP:conf/sigmod/KanthRA02, DBLP:conf/sigmod/FangFNRS08, DBLP:journals/pvldb/ZacharatouDASF17, DBLP:conf/geoinfo/AzevedoZS06, DBLP:conf/icde/SidlauskasCZA18, phtree, geoblocks, deepspace, postgis}). The specific problem of mapping latitude and longitude points to U.S. census blocks \cite{mcelroy2003geocoding,gwinn2018optimal} is available via a variety  of services \cite{TAMUgeo,Tigris}. In addition, our community offered up a wide range of additional suggestions \cite{CGAL,MIT-6.850,JuliaPoly,jones1999first,SCRIP}.

Context plays an important role in selecting fast mapping approaches.  A researcher prototyping an application may prefer a simple approach that can be readily integrated into a high-level rapid prototyping environment.  A developer seeking to maximize performance for a given cost may prefer a more optimized black-box approach that maximizes throughput for a given amount of hardware.  Ultimately, our work focused on two approaches labeled ``simple'' and ``fast''.  The simple approach can be implemented in any scripting language (Matlab/Octave, Python, Julia, R) and is easily integrated and customized to a variety of research goals. This simple approach uses a novel combination of hierarchy, sparse bounding boxes, polygon crossing-number, vectorization, and parallel processing.  The simple approach is compact, does not increase data storage requirements, and is applicable to any country or region.  The fast approach exploits the thread, vector, and memory optimizations that are possible using a low-level language (C++) to achieve optimal performance.  The rest of this paper details these approaches as follows.  A brief description of the location data and the census block polygon data is provided.   The simple mapping approach and the fast mapping approaches are described.  The performance results as a function number  of points, processing threads, and processing cores are given.  Matlab/Octave code for the simple approach is provided in an Appendix and its {\tt inpolygon()} routine is available on GitHub \cite{INPOLY}.  The code for the fast approach is also available on GitHub \cite{FASTMAPPING}.

\section{Data}

\begin{figure}[htb]
  	\centering
    	\includegraphics[width=\columnwidth]{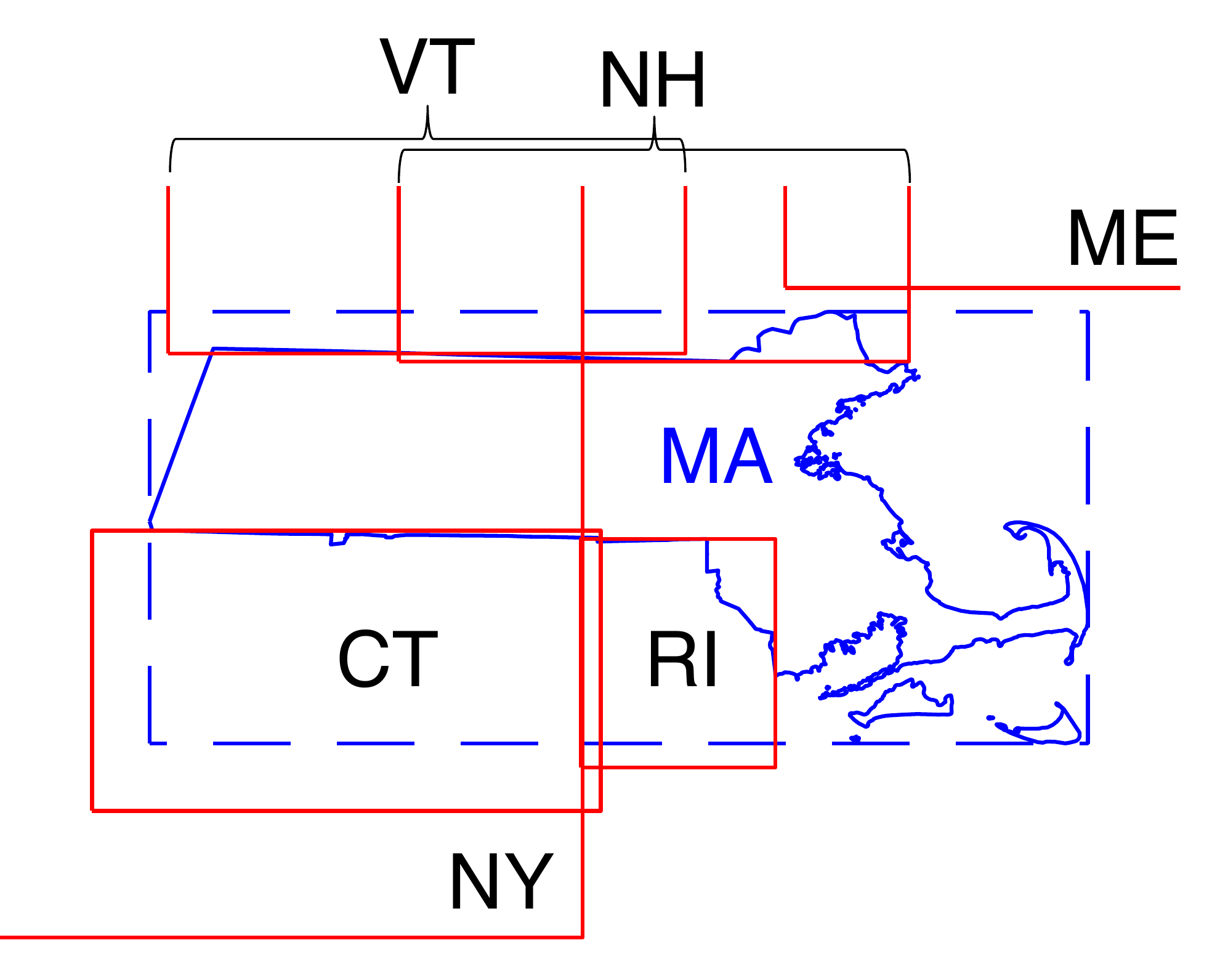}
	\caption{Massachusetts (MA) boundary (solid blue) and bounding box (dashed blue) along with bounding boxes (red) of the neighboring states: New York (NY), Connecticut (CT), Rhode Island (RI), Maine (NH), New Hampshire (NH), and Vermont (VT).}
      	\label{fig:MA-neighbors}
\end{figure}

The location data can be derived from a wide range of sources.   For the purpose of this work, all location data consist of vectors of longitudes, $\bf{x}_{\rm pt}$, and latitudes, $\bf{y}_{\rm pt}$, that are stored in memory as IEEE double-precision floating point numbers.   Demographic boundary data can be any hierarchical representation of the data  from anywhere on the world.   The United States census boundary data consists of 56 states and territories, 3233 counties, and 219831 census block groups \cite{StateCounties,CensusBlocks} (which we refer to as ``census blocks'' for brevity).  These data consist of names, polygon boundaries ($\bf{x}_{\rm poly}$, $\bf{y}_{\rm poly}$), bounding boxes ($\bf{x}_{\rm min}$, $\bf{x}_{\rm max}$, $\bf{y}_{\rm min}$, $\bf{y}_{\rm max}$), and Federal Information Processing Standards (FIPS) codes.  The precise processing goal is to rapidly produce FIPS codes for many latitude and longitude points. Figure~\ref{fig:MA-neighbors} illustrates the polygon boundary for the Commonwealth of Massachusetts (solid blue), its bounding box (dashed blue), and the bounding box of its neighboring states.  Polygon boundaries range from a few points to thousands of points.  Similar polygons and bounding boxes exist for the county and census block group levels in the census hierarchy.

\section{Simple Mapping Approach}

Determining if a set of points is in a polygon is done through calls to an {\tt inpolygon()} function, which has typical computational complexity per point of $O(N_{\rm poly} \log(N_{\rm poly}))$, where $N_{\rm poly}$ is the number of vertices of the polygon.   Primary goals of a fast mapping algorithm are minimizing the number of {\tt inpolygon()} calls and using the fastest {\tt inpolygon()} implementation.  Other goals include minimizing the memory required, minimizing function call overheads, and minimizing the complexity of the algorithm.

Testing points against bounding boxes is significantly faster than calling {\tt inpolygon()}. For example, it is possible to test  all points against state bounding boxes using the expression
\newline

$
{\bf A}_{\rm in} = (\bf{x}_{\rm pt} > \bf{x}_{\rm min}^{\sf T}) ~\&~
                   (\bf{x}_{\rm pt} < \bf{x}_{\rm max}^{\sf T}) ~\&~ 
$

$
~~~~~~~~ (\bf{y}_{\rm pt} > \bf{y}_{\rm min}^{\sf T}) ~\&~
                      (\bf{y}_{\rm pt} < \bf{y}_{\rm max}^{\sf T})
$
\newline

\noindent where
\begin{eqnarray*}
  {\sf T} &=& \text{transpose operation} \\
      \&  &=& \text{logical {\tt and}} \\
       >  &=& \text{logical {\tt greater-than} sparse outer product} \\
       <  &=& \text{logical {\tt less-than} sparse outer product}
\end{eqnarray*}
The resulting $N_{\rm pt} \times N_{\rm state}$ sparse matrix ${\bf A}_{\rm in}$ has the property that
$$
   {\bf A}_{\rm in}(i,j) = 1$$
 if point $i$ is in the bounding box of state $j$.  Let ${\bf 1}$ be an $N_{\rm state} \times 1$ column vector of 1's.  If
$$
  {\bf A}_{\rm in}(i,:) {\bf 1}  = 1
$$
then point $i$ is only in the bounding box of state $j$ and no ${\tt inpolygon()}$ calculations are required.  If
$$
  {\bf A}_{\rm in}(i,:) {\bf 1}  > 1
$$
then ${\tt inpolygon()}$ only needs to be performed on point $i$ for only those states $j$ such that ${\bf A}_{\rm in}(i,j)  = 1$.  Furthermore, all points that  need to be tested against  state $j$ can be determined from the column vector ${\bf A}_{\rm in}(:,j)$, allowing many points to be processed with a single call to ${\tt inpolygon()}$.  The computation of ${\bf A}_{\rm in}$ is dominated by the $>$ and $<$ sparse outer products.  Fortunately, modern sparse matrix libraries that are readily available in high-level programming environments allow these sparse outer products to be vectorized so they can be performed very efficiently on many points at once \cite{KepnerGilbert2011, Mattson2013, kepner2015graphs, kepner2016mathematical, kepner2017enabling, bulucc2017design, davis2017graphblas, Davis-TAMU-2018, kumar2018power9,  cailliau2019redis, Davis-TAMU-2019, Davis:2019:ASG:3375544.3322125}.

To minimize the number of {\tt inpolygon()} calls, the simple algorithm uses bounding box computations along with the natural hierarchy provided by the state, county, and block levels as follows
\begin{enumerate}
  \item State level
  \begin{enumerate}
    \item Test each point against the state bounding boxes
    \item For points in more than one state bounding box, test with {\tt inpolygon()}
  \end{enumerate}
  \item County level
  \begin{enumerate}
    \item Test each point against the county bounding boxes of its state
    \item For points in more than one county bounding box, test with {\tt inpolygon()}
  \end{enumerate}
  \item Block level
  \begin{enumerate}
    \item Test each point against the block bounding boxes of its county
    \item For points in more than one block bounding box, test with {\tt inpolygon()}
  \end{enumerate}
\end{enumerate}
In the above algorithm, {\tt inpolygon()} evaluations only need to be performed on $\sim$20\% of all points ($\sim$0.2 {\tt inpolygon()} evaluations per point).  Furthermore, the points can be organized so that many points can be tested against the same polygon at once.  Vectorizing the {\tt inpolygon()} invocation to process many locations at once reduces the total number of {\tt inpolygon()} calls and minimizes the total call overhead, which can be significant in high-level programming environments.

\subsection{Polygon Crossing-Number}

\begin{figure}[htb]
  	\centering
    	\includegraphics[width=0.75\columnwidth]{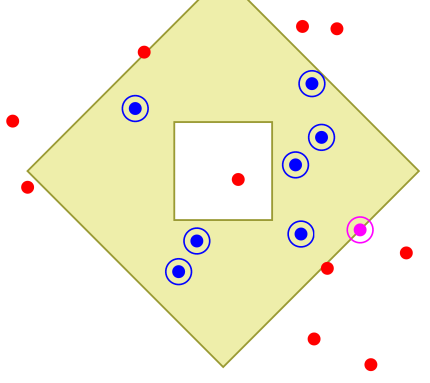}
	\caption{{\tt inpolygon()} returns the inside (blue dots) or outside status (red dots) for a set of locations  and a general polygon (green) embedded in the two-dimensional plane.}
      	\label{fig:InPolygon}
\end{figure}

{\tt inpolygon()} returns the inside/outside status for a set of locations  and a general polygon embedded in the two-dimensional plane (see Figure~\ref{fig:InPolygon})\cite{INPOLY}.  General non-convex and multiply-connected polygonal regions can be handled so the input can be convex or concave, as long as it is not self-intersecting. {\tt inpolygon()} is based on a crossing-number test (introduced in \cite{shimrat1962algorithm}), counting the number of times a line extending from each point past the rightmost region of the polygon intersects with the polygonal boundary.  Points with odd counts are inside the polygon. A simple implementation requires that each of $N_{\rm poly}$ edges  be checked for each of $N_{\rm pt}$ points, leading to an overall complexity of
$$
  O(N_{\rm pt}N_{\rm poly})
$$
   This implementation seeks to improve these bounds. Query points are sorted by y-value and candidate intersection sets are determined via binary search. Given a configuration with an average point-edge overlap of $H$, the overall complexity scales as
$$
  O(N_{\rm poly}H + N_{\rm poly}\log(N_{\rm pt}) + N_{\rm pt}\log(N_{\rm pt}))
$$
where $O(N_{\rm pt}\log(N_{\rm pt}))$ operations are required for the initial sorting, $O(N_{\rm poly} \log(N_{\rm pt}))$ operations are required for the set of binary searches, and $O(N_{\rm poly} H)$ operations are required for the actual intersection tests. $H$ is typically small on average, such that $H << N_{\rm pt}$. Overall, this leads to a fast  complexity for average cases
$$
  O((N_{\rm pt}+N_{\rm poly}) \log(N_{\rm pt}))
$$
Likewise, in well-distributed cases where the number of points overlapping with each edge are on average
$$
  H \approx N_{\rm pt} / N_{\rm poly}
$$
then the overall expected complexity is
$$
  O ((N_{\rm poly}+N_{\rm pt}) \log(N_{\rm pt}) + N_{\rm pt})
$$

\subsection{Data Structure}

The simple approach requires rearranging the state, county, and census block group polygons so that bounding box computations can be rapidly  performed and many points can be tested simultaneously against the same census block group polygon.  At the top country level, this data structure is called {\tt us} and has the following fields containing the state FIPS, bounding box, polygon points, and a state structure for each of 56 state entities

{\footnotesize \begin{verbatim}
us.stateFP: [56x1 double]      
us.stateBB: [56x4 double]
us.stateXY: [56x1 struct]
us.state:   [1x56 struct]
\end{verbatim}}

\noindent For the Commonwealth of Massachusetts these values are

{\footnotesize \begin{verbatim}
us.stateFP(8)   = 25  
us.stateBB(8,:) =
   [-73.5081 -69.9284 41.2380 42.8866]
us.stateXY(8).X = [1x2612 double]
us.stateXY(8).Y = [1x2612 double]
\end{verbatim}}

The {\tt us.state} structure contains the corresponding county FIPS, bounding box, polygon points, and a county structure for each county in each state entity.  For the Commonwealth of Massachusetts there are 14 counties

{\footnotesize \begin{verbatim}
us.state(8).countyFP: [14x1 double]
us.state(8).countyBB: [14x4 double]
us.state(8).countyXY: [14x1 struct]
us.state(8).county:   [1x14 struct]
\end{verbatim}}

\noindent For Middlesex County in the Commonwealth of Massachusetts the values are

{\footnotesize \begin{verbatim}
us.state(8).countyFP(5)   = 17     
us.state(8).countyBB(5)   =
   [-71.8988 -71.0204 42.1568 42.7366]
us.state(8).countyXY(5).X = [1x508 double]
us.state(8).countyXY(5).Y = [1x508 double]
\end{verbatim}}

The {\tt us.state.county} structure contains the corresponding block FIPS, full FIPS, bounding box, and polygon points for each census block group in each county.  The county of Middlesex contains 1133 census block groups

{\footnotesize \begin{verbatim}
us.state(8).county(5).blockFP:   [1133x1 double]
us.state(8).county(5).blockFIPS: [1133x12 char]
us.state(8).county(5).blockBB:   [1133x4 double]
us.state(8).county(5).blockXY:   [1133x1 struct]
\end{verbatim}}

\noindent For one census block group at MIT in Middlesex County these values are

{\footnotesize \begin{verbatim}
us.state(8).county(5).blockFP(488)     = 3531012
us.state(8).county(5).blockFIPS(488,:) = 250173531012
us.state(8).county(5).blockBB(488,:)   =
   [-71.1021 -71.0908 42.3604 42.3660]
us.state(8).county(5).blockXY(488).X   = [1x74 double]
us.state(8).county(5).blockXY(488).Y   = [1x74 double]
\end{verbatim}}

The above structure can be readily computed from standard census data, does not increase the memory requirements of the data, and enables rapid execution of the simple approach to assigning locations to census block group polygons.

\section{Fast Mapping Approach}

Traditional point-in-polygon join algorithms~\cite{jacox2007spatial} follow the filter and refine approach where polygons are prefiltered, typically using bounding boxes, and then refined using computationally expensive point-in-polygon (PIP) tests.

Our fast approach is based on recent work on efficient in-memory point-polygon joins~\cite{kipf2020adaptive,DBLP:conf/icde/KipfLPPB0K18} that makes use of a technique called \emph{true hit filtering}~\cite{brinkhoff1994multi}.  Using \emph{interior} approximations of polygons, this technique allows the identification of join partners already in the filter phase and thus skip expensive refinements in most, if not in all, cases.

\begin{figure}[htb]
  	\centering
    	\includegraphics[width=\columnwidth]{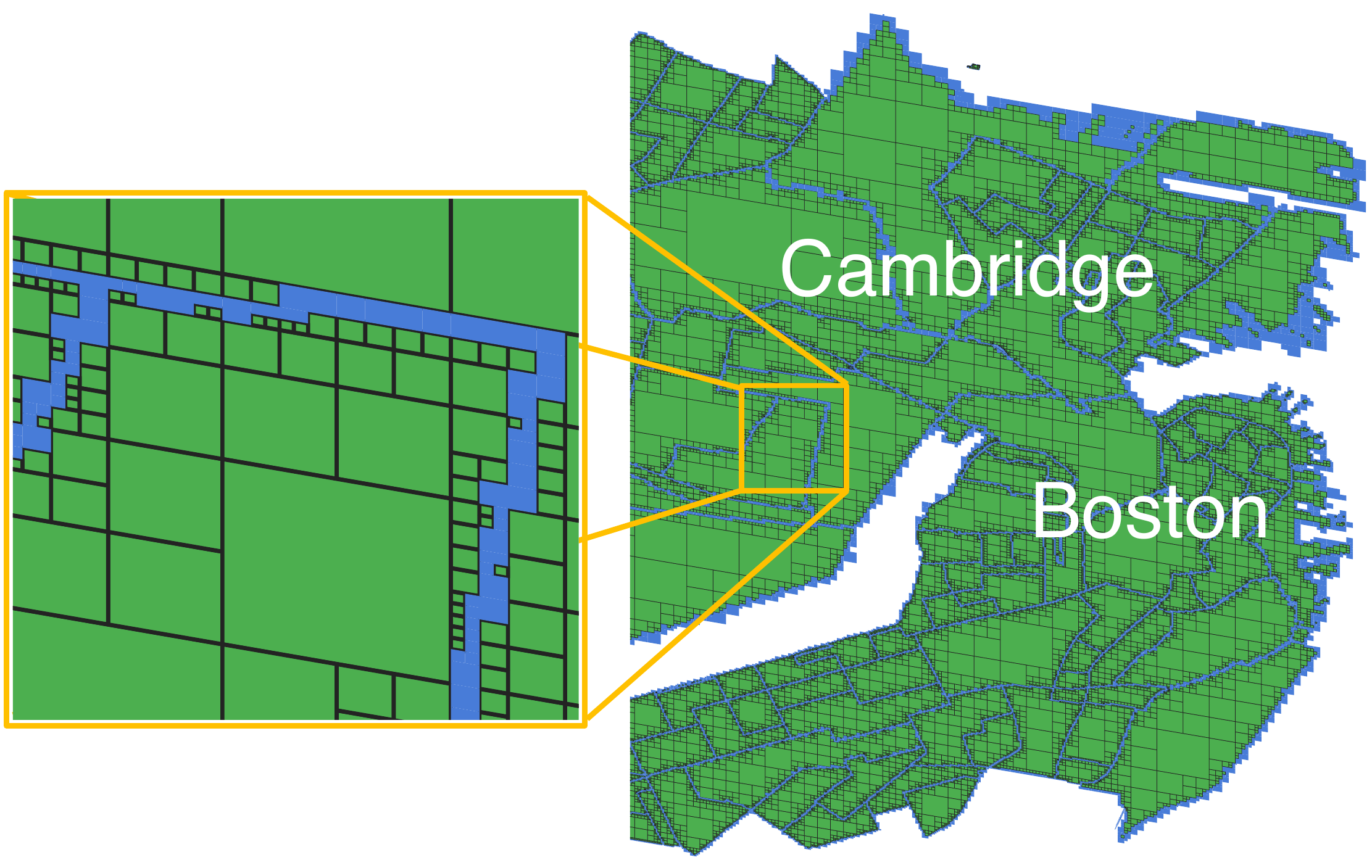}
	\caption{Cambridge and Boston census polygons.  Hierarchical cells correspond to exactly one polygon (green) and one or more polygons (blue).}
      	\label{fig:CambridgeBoston}
\end{figure}

Polygons are approximated by using non-overlapping cells of different sizes with each cell mapping to one or multiple polygons.  A cell either lies fully within a polygon (interior approximation) or covers its boundary.  These approximations are constructed such that most of the polygons' area is covered by interior cells.  Figure~\ref{fig:CambridgeBoston} shows this approximation for the census block groups in the Boston area.  Interior cells are marked in green and boundary cells are marked in blue.  Only in the unlikely event that a query point falls into a boundary cell will a PIP test be performed.

Besides exact results, our fast approach can provide error-bounded approximate results.
When a query point hits a boundary cell in approximate mode, we skip the PIP test and deem the point to be within the polygon.
To control the maximum error (distance) of false positives from their assigned polygon, we subdivide boundary cells into smaller cells such that we guarantee a user-defined precision.
For details on this algorithm, we refer to~\cite{DBLP:conf/icde/KipfLPPB0K18}.

While the exact and the approximate mode significantly reduce the number of or even completely eliminate PIP tests, the drawback is much higher build time and memory consumption compared to our simple approach.

As mentioned before, cells in the polygon approximation do not overlap.
That is, a query point, which is a cell on the most fine-grained grid level, will have at most one matching cell.
All cells are represented as unsigned 64-bit integers obtained by mapping latitude/longitude coordinates to a cube encompassing the earth.
Each cube face is recursively subdivided in a quadtree fashion (each cell is split into four child cells).
For details on this mapping, we refer to documentation on the S2 Geometry library~\cite{S2-lib}.

To efficiently find the matching cell for a query point, we index the cells in a radix tree (trie) data structure.
Compared to the $O(\log n)$ lookup complexity of a B-tree or a binary search on a sorted vector, lookups in a trie are in $O(k)$ with $k$ being the key length (64 in this case).
By using a high tree fanout, the number of node accesses can be further reduced.
For example, in our implementation we index up to four quadtree levels per trie level.
Since it requires two bits to encode a quadtree level, we consume up to 8 bits per trie level (a trie node fanout of $2^8 = 256$).
Another advantage of using a radix tree over a B-tree is that larger cells are indexed closer to the root.
Assuming that larger cells are also more likely to be hit by query points, this indexing yields an additional performance benefit.

\section{Performance}

 Our team has developed a high-productivity scalable platform---the MIT SuperCloud---for providing scientists and engineers the tools they need to analyze large-scale dynamic data \cite{gadepally2018hyperscaling,reuther2018interactive}.  The MIT SuperCloud provides interactive analysis capabilities  accessible from high-level programming environments (Python, Julia, Matlab/Octave) that scale to thousands of processing nodes.   Sparse matrices can be manipulated on the MIT SuperCloud using distributed databases (SciDB and Apache Accumulo), D4M associative arrays \cite{Kepner2012-ch1, kepner2018mathematics, kepner2019d4m}, and now the SuiteSparse GraphBLAS hypersparse matrix library \cite{kepner2020graphblas}.   The benchmarking was performed on the MIT SuperCloud TX-GAIA system (Technology eXperiment - Green Artificial Intelligence Accelerator) consisting of several hundred nodes each with 40 Intel processing cores and 384 GB of RAM with access to a multi-petabyte Lustre filesystem \cite{TX-GAIA-2019}.  The test location data points were drawn from a representative set consistent with mobile device locations.
 
\begin{figure}[ht]
\centering
\includegraphics[width=\columnwidth]{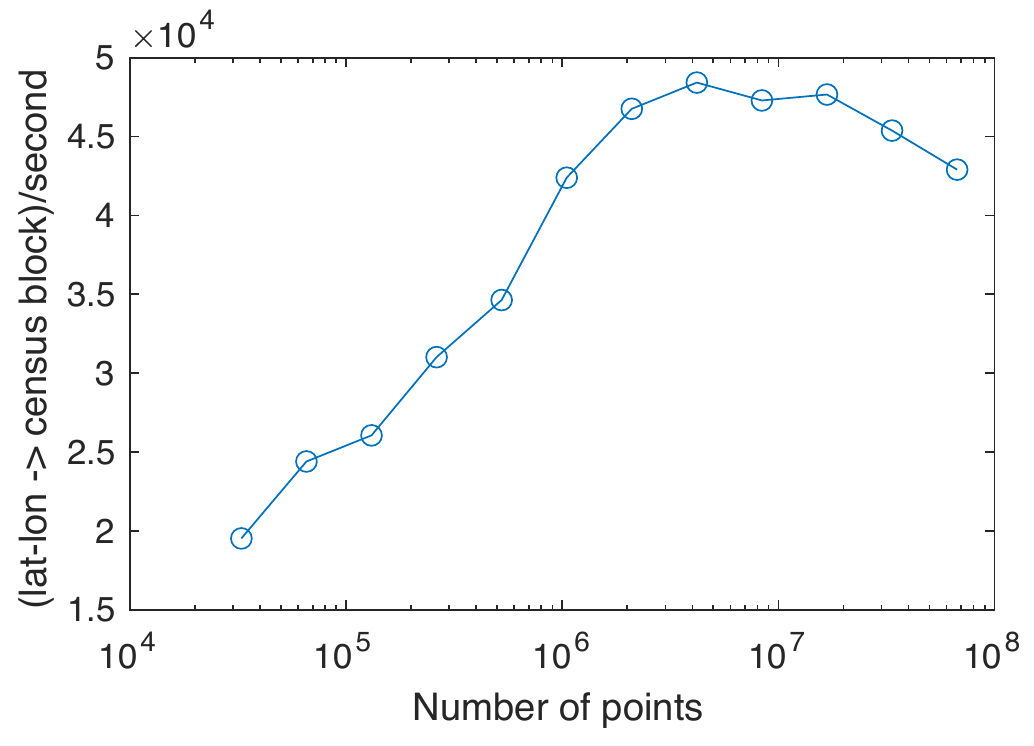}
\caption{Simple Approach: Single-core rate of location (lat-lon) conversion to census blocks as a function of the number of location points processed.}
\label{fig:SimplePointsThroughput}
\end{figure}

\begin{figure}[ht]
\centering
\includegraphics[width=\columnwidth]{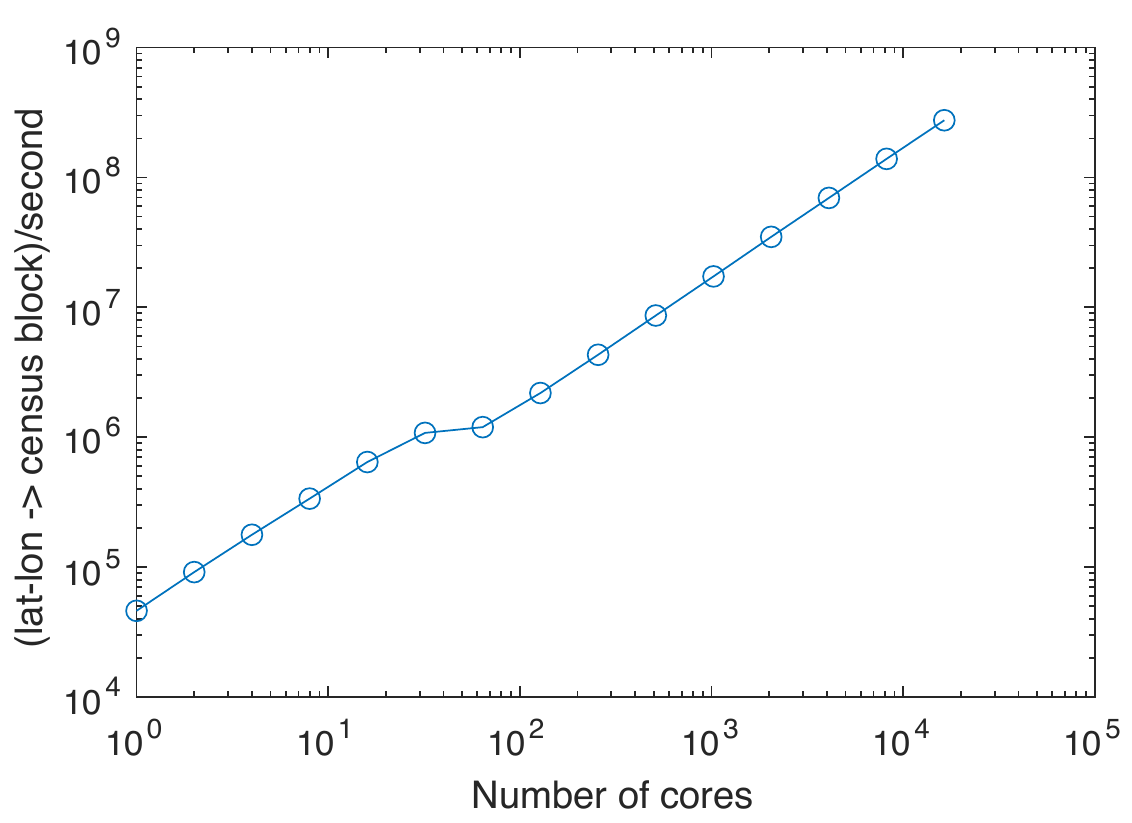}
\caption{Simple Approach: Rate of location (lat-lon) conversion to census blocks as a function of the number of processing cores used.}
\label{fig:SimpleCoresThroughput}
\end{figure}

The simple approach was run in parallel on the MIT SuperCloud using a parallel Matlab library for Matlab/Octave \cite{Kepner2009}.  The single-core performance as a function of the number points is shown in Figure~\ref{fig:SimplePointsThroughput} and peaks between $10^6$ and $10^7$ points at a rate of 45K location assignments per second (lat-lon $\rightarrow$ census blocks).  This peak is most likely due to a balance between the reduced call overhead of using larger numbers of points being offset by less data fitting into higher-speed cache memory.  The simple algorithm is easy to run in parallel as different locations can be processed separately on different processing cores.  Figure~\ref{fig:SimpleCoresThroughput} shows the performance as a function of the number of cores.  The performance on cores 1, 2, 4, 8, 16, and 32 are on a single node.  Running more than 32 instances on a single node did not improve performance. The performance increases linearly up to 16 cores on a node.  Beyond 32 cores, multiple nodes are used and the performance scales linearly, achieving 275M location assignments per second on 256 nodes using 8192 out of 10240 cores.

\begin{figure}[ht]
\centering
\includegraphics[width=\columnwidth]{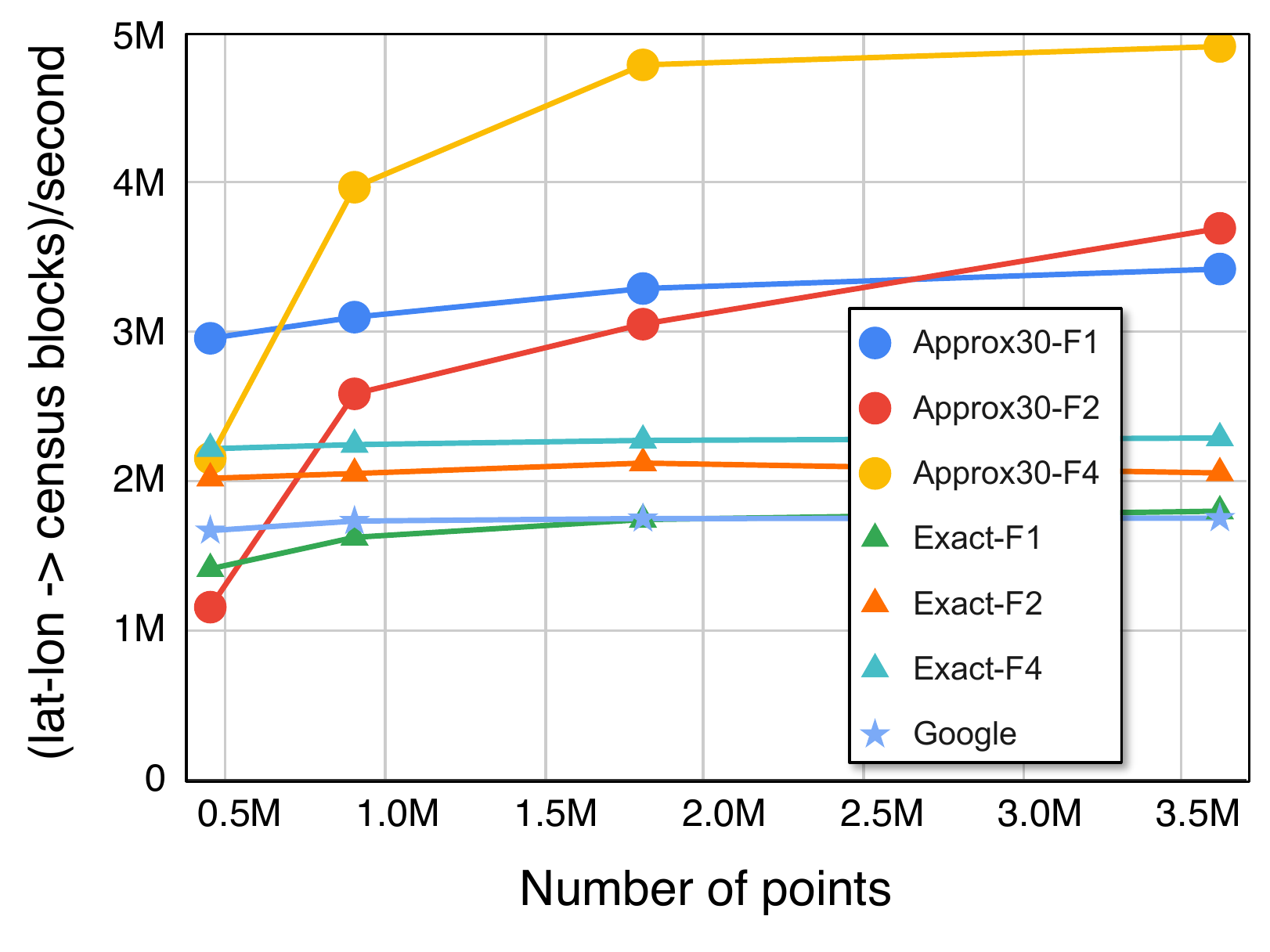}
\caption{Fast Approach: Single-core rate of location (lat-lon) conversion to census blocks as a function of the number of location points processed.}
\label{fig:FastPointsThroughput}
\end{figure}

\begin{figure}[ht]
\centering
\includegraphics[width=\columnwidth]{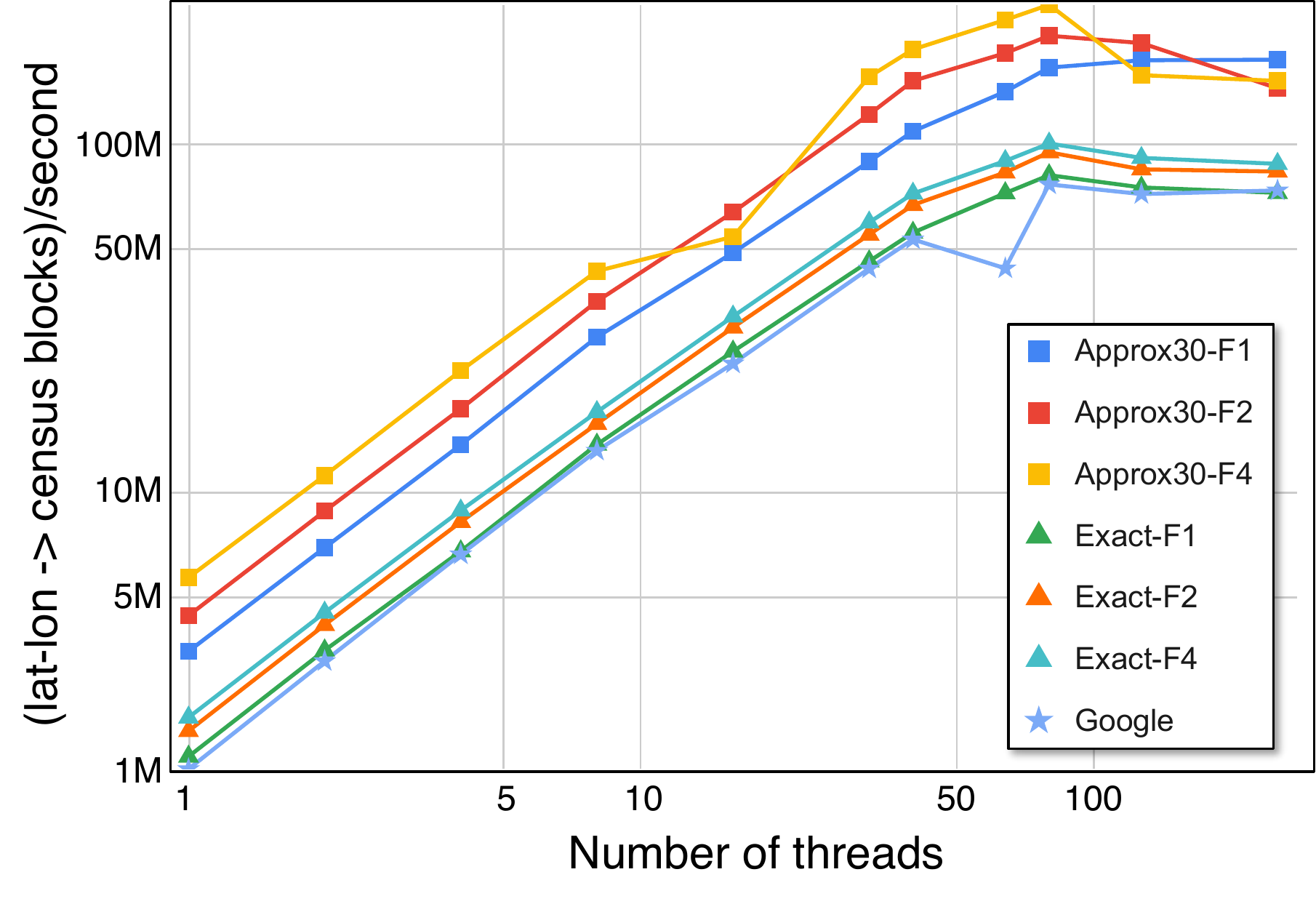}
\caption{Fast Approach: Rate of location (lat-lon) conversion to census blocks as a function of the number of threads used.}
\label{fig:FastThreadsThroughput}
\end{figure}

Both approximate and exact versions of the fast approach were run in parallel on a single MIT SuperCloud node. F1, F2, and F4 denote different trie fanouts with one, two, and four quadtree levels per trie level, respectively. The grid resolution remains the same across these configurations. That is, all of them require the same number of PIP tests (zero in approximate mode). However, with a higher trie fanout, we accelerate the index lookup by trading memory footprint for lookup performance. In addition, measurements were also made using S2ShapeIndex, which is part of the S2 Geometry library~\cite{S2-lib}. In contrast to our fast approach, S2ShapeIndex uses a more coarse-grained grid and uses a B-tree. It provides exact results and does not offer an approximate mode. The single-core performance as a function of the number points is shown in Figure~\ref{fig:FastPointsThroughput} and peaks beyond $10^6$ points at a rate of a few million location assignments per second (lat-lon $\rightarrow$ census blocks). We observe a higher impact of the fanout (F1, F2, and F4) in approximate mode than in exact mode. The reason is that approximate is dominated by index lookups since it does not perform any PIP tests.   The fast algorithm is run in parallel on a single node by enabling more threads.  Figure~\ref{fig:FastThreadsThroughput} shows the performance as a function of the number of threads demonstrating 100M exact location assignments per second.  The performance peaks at 80 threads, which corresponds to the number of physical threads on the node (each core has two hardware threads).

\begin{table}[htp]
\caption{Memory size requirements for fast approaches.}
\begin{center}
\begin{tabular}{lc}
\hline
Approach & Size (GiB) \\
\hline
Approx30-F1 & 39.97 \\
Approx30-F2 & 94.52 \\
Approx30-F4 & 89.37 \\
Exact-F1 & 1.74 \\
Exact-F2 & 2.57 \\
Exact-F4 & 14.59 \\
Google & 1.69 \\
\hline
\end{tabular}
\end{center}
\label{tab:sizes}
\end{table}%

Table~\ref{tab:sizes} shows the space consumption of the different indexes.
Our approximate approach consumes almost 90\,GiB in its most query-efficient configuration.
If space is an issue, one can represent the radix tree in a compact manner that reduces its size to 10\% while still yielding 50\% of its performance~\cite{anneser2020hybrid}.

\section{Conclusion}

Social distancing and contact tracing are pandemic measures that can be made more effective by quickly merging demographic data with location data.  Computing the census block groups of billions of longitude and latitude locations  is computationally challenging and requires performing point-in-polygon evaluations with hundreds of thousands of highly irregular demographic census block group polygons.  This paper offers two approaches for performing these calculations that are appropriate for different contexts.  A simple approach is described that can be implemented in any high-level language to enable a variety of research goals. The simple approach combines hierarchical evaluation, sparse bounding boxes, polygon crossing-number, vectorization, and parallel processing to achieve 100,000,000+ projections per second on 100 servers.  The simple approach is compact, does not increase data storage requirements, and is applicable to any country or region.  A fast approach is also measured that exploits the thread, vector, and memory optimizations that are possible using a low-level language (C++).  The fast  approach achieves similar performance on a single server.  This paper details these approaches with the goal of enabling the broader community to quickly integrate location and demographic data.


\section*{Acknowledgments}
%
%

The authors wish to acknowledge the following individuals for their contributions and support:
Richard Barnes, Enric Boix, Taylor Campbell, Klee Dienes, Yevhenii Diomidov, Dan Feldman, Gregory Galperin, Piotr Indyk, Irwin Jungreis, Ian Katz, Bradley Kuszmaul, Jayson Lynch, Jelani Nelson, Adam Polak, Christopher Rackauckas, Olivia Siegel, Gerald Sussman, Ilya Shlyakhter, Dan Strawser, 
Alan Edelman, Charles Leiserson, Bob Bond, Adam Norige, Dave Martinez, Steve Rejto, Ed Wack, and Marc Zissman.



\bibliographystyle{ieeetr}
\bibliography{FastMapping}
%

\appendices

\section{Simple Approach: Program}
{\tiny \begin{verbatim}
function ptBlockFIPS = findCensusBlock(us,Xpt,Ypt)
  % Assign pionts to census map blocks
  Npt = length(Xpt);

  Nstate = length(us.stateFP);
  Xmin = sparse(us.stateBB(:,1).');
  Xmax = sparse(us.stateBB(:,2).');
  Ymin = sparse(us.stateBB(:,3).');
  Ymax = sparse(us.stateBB(:,4).');

  % Test each point against each state bounding box.
  ptStateBB = (Xpt > Xmin) & (Xpt < Xmax) & ...
              (Ypt > Ymin) & (Ypt < Ymax);

  ptStateMat = ptStateBB;
  ptStateBBcheck = ptStateBB;

  % Leave points in more than one bounding box. 
  ptStateBBcheck(sum(ptStateMat,2) == 1,:) = 0;

  % Resolve points in more than one state bounding box.
  nonEmptyStates = find(sum(ptStateBB,1));
  for i=nonEmptyStates
    % Find points in the state bounding box.
    j = find(ptStateBBcheck(:,i));
    inState = 0;
    if nnz(j)
      inState = inpolygon2(Xpt(j),Ypt(j),us.stateXY(i).X,us.stateXY(i).Y);
      if nnz(inState)
        ptStateMat(j(inState),:) = 0;
        ptStateMat(j(inState),i) = 1;
        ptStateBBcheck(j(inState),:) = 0;
      end
    end
  end

  % Create state vector for each point.
  [ii jj vv] = find(ptStateMat);
  ptState = accumarray(ii,jj,[Npt  1],@max);
  ptStateFP = ptState;
  ptStateFP(ptState > 0) = us.stateFP(ptState(ptState > 0));

  % Assign county to each point.
  ptCounty = zeros(Npt,1);
  ptCountyFP = zeros(Npt,1);
  nonEmptyStates = find(sum(ptStateMat,1));

  for i=nonEmptyStates

    % Get county bounding boxes.
    Xmin = sparse(us.state(i).countyBB(:,1).');
    Xmax = sparse(us.state(i).countyBB(:,2).');
    Ymin = sparse(us.state(i).countyBB(:,3).');
    Ymax = sparse(us.state(i).countyBB(:,4).');

    iState = find(ptState == i);

    if nnz(iState)
      % Test each point against each county bounding box.
      ptCountyBB = (Xpt(iState) > Xmin) & (Xpt(iState) < Xmax) & ...
                   (Ypt(iState) > Ymin) & (Ypt(iState) < Ymax);

      ptCountyMat = ptCountyBB;
      ptCountyBBcheck = ptCountyBB;

      % Leave points in more than one bounding box. 
      ptCountyBBcheck(sum(ptCountyMat,2) == 1,:) = 0;

      Ncounty = length(us.state(i).countyFP);
      nonEmptyCounties = find(sum(ptCountyBB,1));

      for k=nonEmptyCounties
        % Find points in the county bounding box.
        j = find(ptCountyBBcheck(:,k));
        inCounty = 0;
        if nnz(j)
          inCounty = inpolygon2(Xpt(iState(j)),Ypt(iState(j)), ...
            us.state(i).countyXY(k).X,us.state(i).countyXY(k).Y);
          if nnz(inCounty)
            ptCountyMat(j(inCounty),:) = 0;
            ptCountyMat(j(inCounty),k) = 1;
            ptCountyBBcheck(j(inCounty),:) = 0;
          end
        end
      end

      [ii jj vv] = find(ptCountyMat);
      ptCounty(iState) = accumarray(ii,jj,[length(iState)  1],@max);
      ptCountyFP(iState(ptCounty(iState) > 0)) = ...
        us.state(i).countyFP(ptCounty(iState(ptCounty(iState) > 0)));

    end
  end

  % Assign block to each point.
  ptBlock = zeros(Npt,1);
  ptBlockFP = zeros(Npt,1);
  ptBlockFIPS = char(zeros(Npt,12,'int8'));

  for i=nonEmptyStates
    iState = find(ptState == i);
    if nnz(iState)
      Ncounty = length(us.state(i).countyFP);
      ptCounty_iState = ptCounty(iState);
      nonEmptyCounties = unique(ptCounty_iState(ptCounty_iState > 0)).';

      for k=nonEmptyCounties
        iCounty = find(ptCounty_iState == k);
        if nnz(iCounty)

        % Get block bounding boxes.
        Xmin = sparse(us.state(i).county(k).blockBB(:,1).');
        Xmax = sparse(us.state(i).county(k).blockBB(:,2).');
        Ymin = sparse(us.state(i).county(k).blockBB(:,3).');
        Ymax = sparse(us.state(i).county(k).blockBB(:,4).');
        iState_iCounty = iState(iCounty);

        % Test each point against each block bounding box.
        ptBlockBB = (Xpt(iState_iCounty) > Xmin) & (Xpt(iState_iCounty) < Xmax) & ...
                    (Ypt(iState_iCounty) > Ymin) & (Ypt(iState_iCounty) < Ymax);

        ptBlockMat = ptBlockBB;
        ptBlockBBcheck = ptBlockBB;

        % Leave points in more than one bounding box. 
        ptBlockBBcheck(sum(ptBlockMat,2) == 1,:) = 0;
        Nblock = length(us.state(i).county(k).blockFP);
        nonEmptyBlocks = find(sum(ptBlockBB,1));

        for l=nonEmptyBlocks
          j = find(ptBlockBBcheck(:,l));
          inBlock = 0;
          if nnz(j)
            iState_iCounty_j = iState_iCounty(j);
            inBlock = inpolygon2(Xpt(iState_iCounty_j),Ypt(iState_iCounty_j), ...
              us.state(i).county(k).blockXY(l).X,us.state(i).county(k).blockXY(l).Y);
            if nnz(inBlock)
              ptBlockMat(j(inBlock),:) = 0;
              ptBlockMat(j(inBlock),l) = 1;
              ptBlockBBcheck(j(inBlock),:) = 0;
            end
          end
        end

        [ii jj vv] = find(ptBlockMat);
        ptBlock(iState_iCounty) = accumarray(ii,jj,[length(iCounty)  1],@max);
        ptBlockFP(iState_iCounty( ptBlock(iState_iCounty) > 0 )) = ...
          us.state(i).county(k).blockFP(ptBlock(iState_iCounty( ptBlock(iState_iCounty) > 0 )));
        ptBlockFIPS(iState_iCounty( ptBlock(iState_iCounty) > 0 ),:) = ...
          us.state(i).county(k).blockFIPS(ptBlock(iState_iCounty( ptBlock(iState_iCounty) > 0 )),:);

      end
    end
  end
end

function in = inpolygon2(Xpt,Ypt,Xpoly,Ypoly)
  % Format for use with polygon cross-number code.
  nodes = [Xpoly', Ypoly'];
  Npoly = length(Xpoly);
  edges = [(1:Npoly)',(2:(Npoly+1))'];
  edges(Npoly,2) = 1;
  in = inpoly2([Xpt,Ypt],nodes,edges);
end
\end{verbatim}}

\section{Simple Approach: Data Structure}

{\tiny \begin{verbatim}
function us = buildCensusStruct(stateFile,countyFile,blockFile)
  % Build location projection structure from census data.

  S = shaperead(stateFile);    % Read state shape file.
  Nstate = length(S)       
  stateFP = zeros(Nstate,1);
  stateBB = zeros(Nstate,4);
  for i=1:Nstate
    stateFP(i) = str2num(S(i).STATEFP);
    stateBB(i,:) = reshape(S(i).BoundingBox,1,4);
    stateXY(i,1).X = S(i).X;
    stateXY(i,1).Y = S(i).Y;
  end
  us.stateFP = stateFP;
  us.stateBB = stateBB;
  us.stateXY = stateXY;

  C = shaperead(countyFile);   % Read county shape file.
  Ncounty = length(C)
  countyStateFP = zeros(Ncounty,1);
  countyFP = zeros(Ncounty,1);
  countyBB = zeros(Ncounty,4);
  for i=1:Ncounty
    countyStateFP(i) = str2num(C(i).STATEFP);
    countyFP(i) = str2num(C(i).COUNTYFP);
    countyBB(i,:) = reshape(C(i).BoundingBox,1,4);
    countyXY(i,1).X = C(i).X;
    countyXY(i,1).Y = C(i).Y;
  end
  for i=1:Nstate
    istate = find(countyStateFP == stateFP(i));
    us.state(i).countyFP = countyFP(istate);
    us.state(i).countyBB = countyBB(istate,:);
    us.state(i).countyXY = countyXY(istate,:);
  end

  B = shaperead(blockFile);   % Read block shape file.
  Nblock = length(B)
  blockCountyStateFP = zeros(Nblock,1);
  blockCountyFP = zeros(Nblock,1);
  blockFP = zeros(Nblock,1);
  blockFIPS = char(zeros(Nblock,12,'int8'));
  blockBB = zeros(Nblock,4);
  for i=1:Nblock
    blockCountyStateFP(i) = str2num(B(i).STATE_FIPS);
    blockCountyFP(i) = str2num(B(i).CNTY_FIPS);
    blockFP(i) = str2num([B(i).TRACT B(i).BLKGRP]);
    blockFIPS(i,:) = B(i).FIPS;
    blockBB(i,:) = reshape(B(i).BoundingBox,1,4);
    blockXY(i,1).X = B(i).X;
    blockXY(i,1).Y = B(i).Y;
  end
  for i=1:Nstate
   istate = find(blockCountyStateFP == stateFP(i));
   for ii=1:length(us.state(i).countyFP);
    icounty = find(blockCountyFP(istate) == us.state(i).countyFP(ii));
    us.state(i).county(ii).blockFP = blockFP(istate(icounty));
    us.state(i).county(ii).blockFIPS = blockFIPS(istate(icounty),:);
    us.state(i).county(ii).blockBB = blockBB(istate(icounty),:);
    us.state(i).county(ii).blockXY = blockXY(istate(icounty),:);
   end
  end

end
\end{verbatim}}

\end{document}